\documentclass[10pt]{svmult}

\usepackage{makeidx}         % allows index generation
\usepackage{graphicx}        % standard LaTeX graphics tool
\usepackage{mathptmx}             % when including figure files
\usepackage{multicol}        % used for the two-column index
\usepackage[bottom]{footmisc}% places footnotes at page bottom
\usepackage{geometry}
\usepackage{amsmath}
\usepackage{amssymb}
\usepackage{color}
\usepackage{tikz}
\usepackage{helvet}
\usepackage{courier}
\usepackage{footmisc}
\usepackage{algorithm}
\usepackage[noend]{algpseudocode}

\usepackage{graphicx}
\usepackage{caption}
\usepackage{subcaption}

\newtheorem{dfn}{Definition}[section]

\makeindex             % used for the subject index
\begin{document}

\title*{Correlated Equilibria in Wireless Power Control Games}
\author{Sara Berri\inst{1,2}\and
Vineeth Varma\inst{3}\and
Samson Lasaulce\inst{2}\and
Mohammed Said Radjef\inst{1}}
\institute{$^1$ Research Unit LaMOS (Modeling and Optimization of Systems), Faculty of Exact Sciences, University of Bejaia, Bejaia, 06000, Algeria
\texttt{radjefms@gmail.com}
\and $^2$  L2S (CNRS-CentraleSupelec-Univ. Paris Sud), Gif-sur-Yvette, France  \texttt{\{sara.berri, samson.lasaulce\}@l2s.centralesupelec.fr} \and $^3$ CRAN-ENSEM, Nancy, France \texttt{vineeth.satheeskumar-varma@univ-lorraine.fr }}
\maketitle

%Your text goes here. Separate text sections with the standard \LaTeX\
%sectioning commands.
\begin{abstract}\\

In this paper, we consider the problem of wireless power control in an interference channel where transmitters aim to maximize their own benefit. When the individual payoff or utility function is derived from the transmission efficiency and the spent power, previous works typically study the Nash equilibrium of the resulting power control game. We propose to introduce concepts of correlated and communication equilibria from game theory to find efficient solutions (compared to the Nash equilibrium) for this problem. Communication and correlated equilibria are analyzed for the power control game, and we provide algorithms that can achieve these equilibria. Simulation results demonstrate that the correlation is beneficial under some settings, and the players achieve better payoffs.
\end{abstract}

\section{Introduction}
\label{sec:1}

In this work, the notion of correlated equilibrium, which is a generalization of the Nash equilibrium, is applied in the context of power control in wireless networks to determine efficient cooperative strategies.  Power control in wireless networks has been studied using game theory in literature \cite{goodman-pcom-2000,meshkati-jsac-2006} by characterizing the Nash equilibrium. However, as the Nash equilibrium is often inefficient, introducing the concept of correlated equilibrium will improve the players utility.

Several works apply the concept of correlated equilibrium to the wireless communications paradigm, we will now present some of the relevant papers. The papers \cite{paper1,paper2} look at peer-to-peer (P2P) networks, where the behavior of greedy users can degrade the network performance. Here, the authors introduce correlated equilibrium to improve the player utilities. The paper \cite{paper3} study the energy efficiency in ad hoc networks, and propose a cooperative behavior control scheme to determine cooperative strategies, and help non-cooperative players to coordinate their strategies using the correlated equilibrium. An efficient broadcasting strategy in wireless ad hoc networks is proposed in \cite{paper4}, modeling the interaction among nodes as a game, the action set comprises two actions, to forward or drop the received message from the source. To achieve the correlated equilibrium, linear programming, and a distributed learning algorithm based on the regret matching procedure \cite{matching} are used. %In the paper \cite{tansu}, the power control in a code division multiple access (CDMA) system is studied, using the framework of non-cooperative game to obtain control polices. The pricing concept is introduced by defining the cost function as the difference between pricing and the utility function, wherein, the existence and uniqueness of the Nash equilibrium is proved and conditions for convergence of two algorithms are obtained converging to that equilibrium.

%The study is focused on the delay performance analyze of P2P network for the  two-user case and multiple-user case. Different scenarios has been considered, the Nash equilibrium and correlated equilibrium are calculated for each of them. The authors have also simplify the complexity of the correlated equilibria calculation, by dividing the players into two sets. Afterwards, they used the new notations for define the associated new optimization problem.

The main objective of this paper is to propose another equilibrium concept (i.e., correlated equilibrium) in the context of wireless power control games, which allows players to obtain a larger equilibrium set and more efficient points in the presence of a correlation device. Assuming that we can add a correlation device to the game, is it possible to create a mechanism such that the equilibrium payoff set of the obtained game includes payoffs that are not in the initial set (game without correlation mechanism)? We provide answers to this question in this paper.

The key contributions and novelty of our paper are as follows:\begin{enumerate}\item Introduce the concepts of correlated and communication equilibrium to power control in wireless networks.
\item Provide an algorithm to achieve a correlated equilibrium via regret matching.
\item Provide an algorithm to obtain the Pareto-optimal correlated equilibria via linear programming.
\item An extensive numerical study comparing the efficiency of the proposed correlated equilibrium with the standard Nash equilibrium.
\end{enumerate}

%The studied network model is constituted of a source node, a destination node and a set of relays nodes $\mathcal{R}$. The source node broadcasts its data to relays nodes with a some transmit power $p_s^1$. About the relays nodes, they transmit jointly with the source the data to destination, and use the same power $p_s^1=p_{r_{j}}^{2}$ $\forall ~j~\in~\mathcal{R}$. The utility depends only on the energy of the source node and relay nodes. The work is based on the regret matching procedure \cite{matching} to find the correlated equilibrium that improves player's utilities.

\section{System model}
\label{sec:2}

We consider a system comprised of $K \geq 2$ pairs of interfering transmitters and receivers, where the transmitters want to communicate with their respective receivers. The channel gain of the link between Transmitter $i \in \{1,...,K\}$ and Receiver $j \in \{1,...,K\}$ is denoted by $g_{ij} \in \mathcal{G}$ (here $g_{ij}$ are the real valued channel gains), where $\mathcal{G}=\{g^{1},\ldots,g^{N}\}$ represents the alphabet of the possible channel gains. The transmitter $i$ transmits at discrete power level $p_i$ $\in$  $\mathcal{P}_{i}=\{p^{1}_{i},\ldots,p^{M}_{i}\}$, with $p^{1}_{i}=P^{min}_{i}$ and $p^{M}_{i}=P^{max}_{i}$. Note that $\mathcal{G},\mathcal{P}_i \subset \mathbb{R}_{\geq 0}\,  \forall i$.

%%\begin{figure}[h!]
%%\begin{center}
%%\begin{tikzpicture}
%%\node (12) at (3,0.8) {};%{${g_1}$}; %\mathbf
%%\node  (23) at (5,0.8) {};% {$g'_2$};
%%\node  (21) at (3,-0.8) {} ; %{$g_2$};
%%\node  (14) at (1.2,-0.8) {};% {$g'_1$};
%%%\node  (4)  at (0,0) {4};
%%%\node  (1)  at (3,0) {$R$};
%%\node  (2)  at (0,1) {$T_i$};
%%\node  (3)  at (0,-1) {$T_j$};
%%\node  (4)  at (6,1) {$R_i$};%{};
%%\node  (5)  at (6,-1) {$R_j$}; %{};
%%\node  (6)  at (3,1.25) {$g_{ii}$};% {$S_2$};
%%\node  (7)  at (3,-1.25) {$g_{jj}$};
%%\node  (8)  at (2.25,0.5) {$g_{ij}$};
%%\node  (9)  at (3.75,-0.5) {{$g_{ji}$}};
%%\draw [very thick](0,1) circle (0.3cm);
%%\draw [very thick](0,-1) circle (0.3cm);
%%%\draw [very thick](3,0) circle (0.3cm);
%%\draw [very thick](6,1) circle (0.3cm);
%%\draw [very thick](6,-1) circle (0.3cm);
%%\draw[very thick][->,>=latex] (2) to (4);
%%\draw[very thick][->,>=latex] (3) to (5);
%%\draw[very thick][->,>=latex] (2) to (5);
%%\draw[very thick][->,>=latex] (3) to (4);
%%%\draw[very thick][->,>=latex] (1) to (6);
%%\end{tikzpicture}
%%\end{center}
%%\caption{The figure represents the studied scenario, Interference channel model.}\label{fig1}
%%\end{figure}
%%~~~\\

\label{sec:3}
We denote by $\varphi$ a communication efficiency function which measures the packet success rate as a function of the signal to interference and noise ratio (SINR). It is an increasing function and lie in $[0, 1]$, and is identical for all the users. Let $\mathcal{A}_i$ denote a finite discrete set of actions that can be taken by player $i$. In a power control game, this action corresponds to the wireless signal power used by the transmitter $i$. The SINR at receiver $i \in \mathcal{K}$  writes as:
 \begin{equation}\label{sinr}
    \mathrm{SINR}_i=\frac{a_ig_{ii}}{\sigma^{2}+\sum_{j\neq i} a_jg_{ji}}.
  \end{equation}
   where $a_i \in \mathcal{A_i}$ is the power of the transmitter $i$ and $\sigma^{2}$ is the noise power.

Using these notations, the power control game denoted by $\mathcal{G}$, is defined in its normal form as follows.
\begin{dfn}
A power control game is a triplet:
\begin{equation}\label{game}
    \mathcal{G}=\{\mathcal{K}, \{\mathcal{A}_i\}_{i\in\mathcal{K}},\{u_i\}_{i\in \mathcal{K}}\};
\end{equation}
where:
\begin{enumerate}
  \item  $\mathcal{K}=\{1,\ldots,K\}$ is the set of players.% We note the player set $\{1, \ldots,k\}$.
  \item $\mathcal{A}_{i}=\{a^{1}_{i},\ldots,a^{M}_{i}\}$, is the corresponding power level set of the player $i$ $\in \{1,\ldots,K\}$; with $a^m$ sorted such that $a^m < a^{m+1} m=\overline{1,M}$, $a^{1}_{i}=P^{\min}_{i}$ and $a^{M}_{i}=P^{\max}_{i}$, are respectively the minimum and maximum transmitting power of player $i$, $M\geq1$.%,  chooses a transmitting power
  %\item \textbf{Utility:} each player $i$ has a utility function denoted $u_i$.
  \item $u_1,\ldots,u_K$ are the utility functions of the $K$ players for a combination of choices $a=(a_1,\ldots,a_K)=(a_i,a_{-i})$, where $a_{-i}=(a_1,\ldots,a_{i-1},a_{i+1},\ldots,a_K)$ denotes the power levels of all other players except player $i$. For each player $i$, the utility function $u_i$ depends on the success of its transmission, which is a function of all players' actions through $a$, and on the energy spent in transmission $a_i$.  Mathematically, the players' utilities are defined by the following formula (\ref{uti}):
  %Let $r_i$ the reward of a transmitter $i$ in the transmission. It takes the following form:
  \begin{equation}\label{uti}
    u_i(a_1,\ldots,a_K)=\varphi(\mathrm{SINR}_i)-\alpha a_i,
  \end{equation}
 The parameter $\alpha>0$, is introduced to weigh the energy cost.\\
\end{enumerate}
\end{dfn}

Note that this kind of payoff, different from the traditional energy efficiency (the ratio of the data rate to the power) has been studied in literature \cite{tansu}, and is relevant when the payoff corresponds to the profit (in terms of money) for each step. In the following section, we introduce the problem of correlated equilibrium from the power control game, where an observer is added to help the players to correlate their actions.

  %\item The cost of the transmitter $i$ is noted $c_i$, it depends on the chosen action. The parameter $\alpha>0$, is introduced to measure the importance of the energy consumption.%It equals to:
    %\begin{equation}\label{eq2}
%    c_i=\alpha a_i,
%  \end{equation}
  %where: $\alpha>0$, and $a_i \in \{0,P\}$. \textcolor[rgb]{1.00,0.00,0.00}{We assume $r>c$.}
  % \begin{equation}\label{eq5}
 %  u_{i}(a_{i},a_{-i})= \mathbb{E}_g\varphi(SINR_i)-\alpha a_i,
%\end{equation}
%where: $SINR_i=\frac{a_ig_i}{\sigma^{2}+\sum_{j\neq i} a_jg_j}$. $\sigma^{2}$ is the noise variance. $\varphi$ is an efficiency function. $g_i$ is the channel gain of the transmitter $i$, and $\alpha>0$.%$p_i=P^{max}$

\section{Problem formulation}

In the proposed game (\ref{game}), the users are modeled as rational players, which means they are expected to
choose actions from the possible choices to maximize their utilities. An important concept to characterize the outcome of the game is the Nash equilibrium, which states that every player will select an action which maximizes its utility given the actions of every other player. It corresponds to an action profile from which no player has interest in changing unilaterally its action. However, it is well known that the Nash equilibrium does not always lead to the best performance for players \cite{goodman-pcom-2000,meshkati-jsac-2006}.

Therefore, other concepts to reach such a more efficient equilibrium need to be investigated. In this paper, we will study the concepts of correlated equilibrium and communication equilibrium, introduced by Aumann \cite{aum}, and Forges \cite{forges1}, respectively. These concepts allow players to use an external mediator which provides each of them information about the action to be played in the game. Such a coordination scheme between players may sustain some equilibrium payoffs that are not achievable by an equilibrium without it (Nash equilibrium). The same conclusions hold for the sub-game perfect correlated equilibrium in repeated games, \cite{lehrer}. Moreover, the correlated equilibrium is simpler to compute than the Nash equilibrium \cite{complex}. In the following sub-section, we define the extended game, including an outside observer, and the correlated equilibrium concept.

\subsection{Power control games using a correlated device}
\label{sec:4}
The concept of correlated equilibrium is developed by considering an extended game that includes an outside observer (mediator), which provides each user with a private recommendation regarding which action to perform. The recommendations are chosen according to a probability distribution over the set of action profiles. This probability distribution is called a correlated equilibrium, if the action profile in which all players follow the observer's recommendations is a Nash equilibrium of the extended game.
\begin{dfn}\label{dfn1}
%A canonical correlated equilibrium is induced by a canonical correlation device, that is a correlation device in which $\Omega=\mathcal{A}=\prod_{i\in \mathcal{K}} \mathcal{A}_i,$ and the partition ${\mathcal{P}}_{i}$  is generated by $\mathcal{A}_i$. Thus, a probability distribution $p$ over $\mathcal{A}$ is a canonical correlated equilibrium if and only if. %For every player

Define $p$ as a joint probability distribution on the action profile set $\mathcal{A}=\prod_{i\in \mathcal{K}} \mathcal{A}_i$. The distribution $p$ is a correlated equilibrium if and only if for every player $i$ $\in$ $\mathcal{K}$:
\begin{equation}\label{15}
 \sum_{a_{-i}\in \mathcal{A}_{-i}}p (a_i,a_{-i})u_i(a_i,a_{-i})\geq  \sum_{a_{-i}\in \mathcal{A}_{-i}}p(a_i,a_{-i})u_i(a'_i,a_{-i}), \forall a_i, a'_i\in \mathcal{A}_i.
\end{equation}
\end{dfn}

Inequality (\ref{15}) means that for each user $i$, choosing power level $a'_i$ while it received recommendation to choose the power $a_i$, does not provide a higher expected utility. Thus, it is in the best interest for the users to follow the recommended action. The set of correlated equilibria is nonempty, closed and convex in every finite game \cite{matching}, and it can include distributions that are not in the convex hull of the Nash equilibria distributions. Indeed, Nash equilibrium is a special case of correlated equilibria, where $p(a_i,a_{-i})$ corresponds to the product of each individual probability. Thus, the set of Nash equilibrium points is a subset of the set of the correlated equilibrium points.

The correlated equilibrium considers the ability of users to coordinate actions, and computes the optimality by the joint distribution, so it provides a better solution compared with the non-cooperative Nash equilibrium, where each user acts in isolation. Further, the correlation can provide important insights, when we face the problem of equilibrium selection in games admitting multiple equilibria, which could be the case of the proposed game in some setups.

Now, if the problem is to add a mediator to the game  and enable players to coordinate their actions, but also assume that the mediator receives information from the players and afterwards gives them recommendations, we could construct another kind of mechanism, that makes the information coming from players as inputs, and uses them to find the suitable outputs that correspond to the recommendations. In this scenario, the mechanism is also created such that the players have no interest to deviate from the recommendations, which is the communication equilibrium concept. This information exchange could bring performance improvement with respect to the case where players coordinate their actions without reporting any information to the mediator. In the following Section \ref{sec:5}, we present how we could apply this concept to the power control game (\ref{game}).%, by defining its Bayesian version.

\subsection{Communication equilibrium in power control games}
\label{sec:5}
Here, we assume that the mediator collects information from the players before making them recommendations. In the studied power control game, we assume the players have type sets represented by $\mathcal{T}_i=\{(g_{ii}^{1},\ldots,g_{Ki}^{1}),\ldots,(g_{ii}^{N},\ldots,g_{Ki}^{N})\}$. In the case of a wireless channel, the channel gains are randomly chosen following a certain (known) probability distribution. Therefore, the 'type' for each player is chosen randomly according to the given probability distribution over each player's type set. Each player $i$ sends information about his type, i.e., $t_i \in  \mathcal{T}_i$ to the mediator (the player might lie if it brings some gain). Thus, a communication device consists of a system $p$ of probability distributions $p = p(.|t)_{t\in \mathcal{T}=\prod_{i\in \mathcal{K}} \mathcal{T}_i}$ over $\mathcal{A}=\prod_{i\in \mathcal{K}} \mathcal{A}_i$. The interpretation is that every player $i\in \mathcal{K}$ reports its type $t_i=(g_{ii},g_{ji})_{\forall j\neq i}$ to the mediator, which privately recommends $a_i$ according to $p=p(.|t)$.  The system $p$ defines a communication equilibrium if none of the players can gain by unilaterally lying on its type or by deviating from the recommended action.
   \begin{dfn}\label{dfn11} %\textbf{Communication equilibrium}\\
    The system $p$ defines a communication equilibrium if:
\begin{multline}\label{11}
  \sum_{t_{-i}\in \mathcal{T}_{-i}}q(t_{-i}|t_i)\sum_{a\in \mathcal{A}} p(a|t)u_i(t,a)\geq  \sum_{t_{-i}\in \mathcal{T}_{-i}}q(t_{-i}|t_i)\sum_{a\in \mathcal{A}} p(a|t'_i,t_{-i})u_i(t,a'_i,a_{-i}) \forall i\in \mathcal{K},~ \forall t_i, t'_i \in \mathcal{T}_i,\forall a'_i\in \mathcal{A}_i.
\end{multline}
\end{dfn}
where $q(t_{-i}|t_i)=\frac{q(t)}{\sum_{t_i\in T_i}q(t_i,t_{-i})}$, is the subjective probability assigned to the event $t_{-i}$, that is the actual profile of the other players' types, if $t_i$  is the type of $i$. The definition \ref{dfn11} implies that player $i$ does not get a higher expected utility if it lies about its true type $t_i$ and reports $t'_i$, or if it plays another action $a'_i$ instead of recommended action $a_i$. In the following section, we propose some techniques to achieve correlated and communication equilibria.

 \section{Implementation of communication and correlated equilibria}
 \label{sec:6}
 The sets of correlated equilibria and communication equilibria are the subsets defined by the intersection of the half-spaces given by the inequalities (\ref{15}) and (\ref{11}), respectively. In this section, we investigate methods to obtain correlated and communication equilibria.

\subsection{Linear programming method}

In this paper, among the multiple correlated equilibria, we consider those which provide the highest social welfare. Thus, the problem boils down to computing a correlated equilibrium which maximizes the sum of the players'
expected utilities. In order to characterize these equilibria, we propose to use a linear programming method in which the optimization problem of the power control game can be formulated as follows:
      %\begin{equation}\label{OF}
%            % \sum_{i \in N} \sum_{a_i,a_{-i}} p(a_i,a_{-i})u_i(a_i,a_{-i}).
%             \max \sum_{a \in \mathcal{A}} p(a)\sum_{i \in \mathcal{K}} u_i(a).
%      \end{equation}
%      subject to constraint (\ref{15})
%      \begin{equation}\label{121}
% \sum_{a_{-i}\in \mathcal{A}_{-i}}p (a_i,a_{-i})u_i(a_i,a_{-i})\geq  \sum_{a_{-i}\in \mathcal{A}_{-i}}p(a_i,a_{-i})u_i(a'_i,a_{-i}), \forall i~\in~\mathcal{K}~and~ \forall a_i, a'_i\in \mathcal{A}_i.
%\end{equation}
%      and %for all player $i~\in~\mathcal{K}$ and actions $a_i, a'_i\in \mathcal{A}_i$.
%      \begin{equation}\label{12}
%        \sum_{a_i,a_{-i}} p(a_i,a_{-i})=1.
%      \end{equation}
%      \begin{equation}\label{13}
%         p(a_i,a_{-i})\geq 0.
      %\end{equation}
      %\item 
    %\end{itemize}

%%%%%%%%%%%%%%%%% The new formulation%%%%%%%%%%%%%
\begin{equation}\label{li1}
\begin{array}{llllll}
\max c^Tx &  \\
   A_ix\geq0 ~~\forall i \in \mathcal{K}\\
    \sum_{j=1}^{M^K}x_j=1;  \\
 0 \leq x_j\leq 1 &\forall j=\{1,\ldots,M^K\}   \\
     \end{array}
     \end{equation}
Where $x^T=(p(a_1^1,\ldots,a_K^1),\ldots,p(a_1^M,\ldots,a_K^M))$; $c^T=(\sum_{i\in \mathcal{K}}u_i(a_1^1,\ldots,a_K^1),\ldots,\sum_{i \in \mathcal{K}}u_i(a_1^M,\ldots,a_K^M))$.
\tiny \begin{equation}\label{cons1}
    A_i=\bordermatrix{
&& &\cr
& (u_i(a_i^1,a_{-i}^1)-u_i(a_i^2,a_{-i}^1))
& \ldots &(u_i(a_i^1,a_{-i}^M)-u_i(a_i^2,a_{-i}^M))&0&\ldots&0&0&\ldots&0 \cr & \vdots& \vdots& \vdots& \vdots&\vdots&\vdots&\vdots&\vdots&\vdots&\cr & (u_i(a_i^1,a_{-i}^1)-u_i(a_i^M,a_{-i}^1))
& \ldots &(u_i(a_i^1,a_{-i}^M)-u_i(a_i^M,a_{-i}^M))&0&\ldots&0&0&\ldots&0 \cr
&0
& \ldots &0& (u_i(a_i^2,a_{-i}^1)-u_i(a_i^1,a_{-i}^1))&\ldots&(u_i(a_i^2,a_{-i}^M)-u_i(a_i^1,a_{-i}^M))&0&\ldots&0 \cr &0
& \ldots &0& (u_i(a_i^2,a_{-i}^1)-u_i(a_i^3,a_{-i}^1))&\ldots&(u_i(a_i^2,a_{-i}^M)-u_i(a_i^3,a_{-i}^M))&0&\ldots&0 \cr & \vdots& \vdots& \vdots& \vdots&\vdots&\vdots&\vdots&\vdots&\vdots&\cr &0
& \ldots &0&(u_i(a_i^2,a_{-1}^1)-u_i(a_i^M,a_{-i}^1))&\ldots&(u_i(a_i^2,a_{-1}^M)-u_i(a_i^M,a_{-i}^M))&0&\ldots&0 \cr & \vdots& \vdots& \vdots& \vdots&\vdots&\vdots&\vdots&\vdots&\vdots&\cr &0
& \ldots &0& 0&\ldots&0&(u_i(a_i^M,a_{-i}^1)-u_i(a_i^{1},a_{-i}^1))&\ldots&(u_i(a_i^M,a_{-i}^M)-u_i(a_i^{1},a_{-i}^M)) \cr & \vdots& \vdots& \vdots& \vdots&\vdots&\vdots&\vdots&\vdots&\vdots&\cr &0
& \ldots &0&0&\ldots&0&(u_i(a_i^M,a_{-i}^1)-u_i(a_i^{M-1},a_{-i}^1))&\ldots&(u_i(a_i^M,a_{-i}^{M})-u_i(a_i^{M},a_{-i}^{M-1}))}.
   \end{equation} \normalsize

     $p(a_i,a_{-i})$ is the probability that the action profile $(a_i,a_{-i})$ is chosen. Thus, a distribution $x^*$ is said to be optimal correlated equilibrium if it is solution of the linear program (\ref{li1}).%satisfies the constraints (\ref{121})-(\ref{13}) and provides the maximum of (\ref{OF}).
\begin{algorithm}
\caption{Algorithm leading to the optimal correlated equilibrium in power control game}\label{alg2}
\begin{algorithmic}[1]
\State  In the beginning, the mediator chooses
a power profile $(a_i,a_{-i})$ $\in$ $A$ according to $p^*$, that is obtained solving the described linear program (\ref{li1}) using an appropriate method, $p^*=x^*$.
\State The mediator informs each user $i$ of the power to choose $a_i$.
%%profile and the sequence of inputs and outputs., and construct the corresponding  and performs lotteries by using methods to solve linear program

\end{algorithmic}
\end{algorithm}
 %\subsubsection{Description of the game}
% \begin{enumerate}
%   \item  Before the game starts the mediator chooses
%a point $(a_i,a_{-i})$ $\in$ $A$ according to $p^*$, that is obtained using linear programming methods.
%   \item The mediator informs each user $i$ of $a_i$.

% \end{enumerate}
    In the same manner, we can characterize the optimal communication equilibrium. The solution could be obtained by solving the following optimization problem:
   \begin{equation}\label{11b}
  \max \sum_{a\in \mathcal{A} ~~t\in \mathcal{T}}q(t) p(a|t) \sum_{i\in \mathcal{K}}u_i(t,a).
\end{equation}
      subject to constraint (\ref{11})
      \begin{multline}\label{11a}
  \sum_{t_{-i}\in \mathcal{T}_{-i}}q(t_{-i}|t_i)\sum_{a\in \mathcal{A}} p(a|t)u_i(t,a)\geq  \sum_{t_{-i}\in \mathcal{T}_{-i}}q(t_{-i}|t_i)\sum_{a\in \mathcal{A}} p(a|t'_i,t_{-i})u_i(t,a'_i,a_{-i}) \forall i\in \mathcal{N},~ \forall t_i, t'_i \in \mathcal{T}_i,\forall a'_i\in \mathcal{A}_i.
\end{multline}
and for all $t\in \mathcal{T}$.
       %$i~\in~N$ and actions $a_i, a'_i\in A_i$.
      \begin{equation}\label{12a}
        \sum_{a\in \mathcal{A}} p(a|t)=1.
      \end{equation}
      \begin{equation}\label{13a}
         p(a|t)\geq 0.
      \end{equation}

      %Thus, we can use linear programming methods to obtain the optimal correlated and communication equilibria.\\

       In the following we summarize the different steps to reach the optimal communication equilibrium.
      \begin{algorithm}
\caption{Algorithm leading to the optimal communication equilibrium in power control game}\label{alg}
\begin{algorithmic}[1]
\State The mediator simulates the sequence of reports that the users could send, that correspond to channel gain profiles, and the power profiles that could be received by the users given a channel gain profile.
\State Using a method to solve the linear program constituted by the objective function (\ref{11b}) and the constraints (\ref{11a})-(\ref{13a}), to find the optimal probability distributions $p(.|t)$ for all type profile $t$.
%%profile and the sequence of inputs and outputs., and construct the corresponding  and performs lotteries by using methods to solve linear program
\State For each player $i$, the Nature randomly chooses type, that corresponds to the channel gain profile $(g_{ii},\ldots,g_{Ki})$, according to a given probability distribution over the type set $T_i$.
\State Each player $i$ reports its type, $t_i$, to the mediator.
\State The mediator performs lotteries according to the received type profile, and sends private recommendations to the players, that correspond to the transmitting powers.
\end{algorithmic}
\end{algorithm}

      %the mediator first
%simulates the sequence of signals and reports (inputs) that would have been sent by
%the players and the sequence of messages (outputs) that would have been received by
%the players given the type profile under the original equilibrium. Then, he computes
%the actions that would have been chosen by the players as a function of the type
%%profile and the sequence of inputs and outputs. Finally, he privately recommends each player to choose the associated action. Clearly, if a player has an incentive to
%deviate from the recommendation of the mediator, then the strategy profile of the
%original communication game was not an equilibrium.

However, with the linear programming method, the computation complexity grows exponentially with the number of users and actions since an increase in the number of users and actions, results an increase in the number of constraints. There exists a distributed learning approach, i.e., regret matching \cite{matching} to achieve a correlated equilibrium. However, regret matching does not ensure Pareto optimality of the given correlated equilibrium as it converges to an arbitrary correlated equilibrium, whereas the linear program as defined in (\ref{li1}) gives a correlated equilibrium that maximizes the social welfare. In the following, we present the distributed learning approach.

\subsection{Regret matching procedure}
A game procedure is proposed in \cite{matching}, called 'regret-matching'. In which the players measure the regret for not choosing other actions in the past, and change their current action with probabilities that are proportional to these measures. Thus, the game is played with probability distribution over the action set. The
details of the regret-matching algorithm \cite{matching} are shown in the Algorithm \ref{alg3}.
\begin{algorithm}
\caption{Regret matching algorithm for power control game}\label{alg3}
\begin{algorithmic}[1]
\State For any two distinct power levels $a'_i\neq a_i$ $\in$  $A_i$ calculate the average regret of user $i$ at time $t$ for not choosing $a'_i$ as:
\begin{equation}\label{regret}
    R^{t}_{i}(a_{i},a'_{i})=\max \{\frac{1}{t}\sum_{\tau \leq t}[u_i(a'_i,a_{-i}^{\tau})-u_i(a_i^{\tau})],0\}.
\end{equation}
\State Let $a_i$ $\in$ $A_i$, the last power chosen by user $i$, $a^{t}_i=a_i$. Then the probability distribution over the
power levels for the next period, is defined as
$$
\left\{
  \begin{array}{ll}
    p_i^{t+1}(a'_i)=\frac{1}{\mu}R^{t}_{i}(a_{i},a'_{i}), & \hbox{$\forall a'_i\neq a_i$;} \\
    p_i^{t+1}(a_i)=1-\sum_{a'_i \in A_i a'_i\neq a_i} p_i^{t+1}(a'_i), &
  \end{array}
\right.
 $$
where $\mu$ is an enough large constant.
\State For every $t$, the empirical distribution of the power profile $a$ is:
\begin{equation}\label{proba}
    p^{t}(a)=\frac{1}{t}|\{\tau \leq t: a^{\tau}=a\}|.
\end{equation}
where $|\{\tau \leq t: a^{\tau}=a\}|$ is the number of times the power profile $a$ has been chosen in the periods before $t$.
\end{algorithmic}
\end{algorithm}

It is shown in \cite{matching} that if the players implement  Algorithm \ref{alg3}, the empirical distribution (\ref{proba}) converges to an arbitrary correlated equilibrium of the game if it is not unique. The obtained correlated equilibrium by applying this procedure is not always Pareto-optimal (but the procedure can achieve the PO equilibrium). However, the one provided by the linear programming method is Pareto-optimal.

 \section{Numerical Results and Analysis}
 \label{sec:7}
 In this section, we present numerical results. The simulation setup is as follows. The number of Transmitter and Receiver pairs is equal to 2,  $\mathcal{K}=\{1,2\}$. Set of possible powers, that is the action set $\mathcal{A}_i$, $\forall i \in  \mathcal{K}$: $M=25$, $P_i^{\min}(\mathrm{dB}) = -20$, $P_i^{\max}(\mathrm{dB})=+20$. The presented results correspond to the expected values over different values of $({g}_{11},{g}_{12},{g}_{22},{g}_{21})$ that lie in a discrete set $\mathcal{G}_i=\{g^{1}_{i},\ldots,g^{N}_{i}\}$, with $g^{1}_{i}=g^{\min}_{i}$, $g^{N}_{i}=g^{\max}_{i}$, $\forall \in i \mathcal{K}$: $N=10$, $g_i^{\min} = 0.01$, $g_i^{\max}=3$, the channel gain increment equals  $\frac{3-0.01}{10}$. The means of the channel gains are given by: $(\bar{g}_{11},\bar{g}_{12},\bar{g}_{22},\bar{g}_{21})=(1,1,1,1)$. The communication efficiency function is:  $\varphi(x) = (1-e^{-x})^L$, $L$ being the number of symbols per packet (see e.g., \cite{goodman-pcom-2000}\cite{meshkati-jsac-2006}\cite{lasaulce-twc-2009}). In most of the simulations provided we take $L=100$. The aforementioned parameters are assumed, otherwise they are explicitly mentioned in figures.

%\begin{figure}[h]
%	\centering
%	\begin{subfigure}[t]{.48 \linewidth}
%		\centering
%		 \includegraphics[width=7cm,height=5cm]{VSPvar26-08.pdf}
%		\caption{The gain of correlated and Nash equilibria when compared to expected payoffs obtained by the global optimum for sum-utility, against different number of actions $|A_i|$. \textcolor[rgb]{1.00,0.00,0.00}{The different sets $|A_i|$ are constructed in an increasing manner.}} \label{figA}
%	\end{subfigure}
%\hspace{0.01\linewidth}
%	\begin{subfigure}[t]{.48 \linewidth}
%		\centering
%		 \includegraphics[width=7cm,height=5cm]{VS-SIR.pdf}
%		\caption{The gain of correlated and Nash equilibria when compared to expected payoffs obtained by the global optimum for sum-utility, against different values of signal to interference.}  \label{fig2}
%	\end{subfigure}
%	\caption{}
%\end{figure}

%
%In figure \ref{fig21}, when the number of possible actions $|A_i|$ equals to $2$, the global optimum, correlated and Nash equilibria provide the same value of the sum-utility, due to the existence of dominated action that corresponds to all the different solutions. In such cases, correlation could help to select an equilibrium if the game admits several equilibria.
\begin{figure}[h]
	\centering
	\begin{subfigure}[t]{.48 \linewidth}
		\centering
		 \includegraphics[width=7cm,height=5cm]{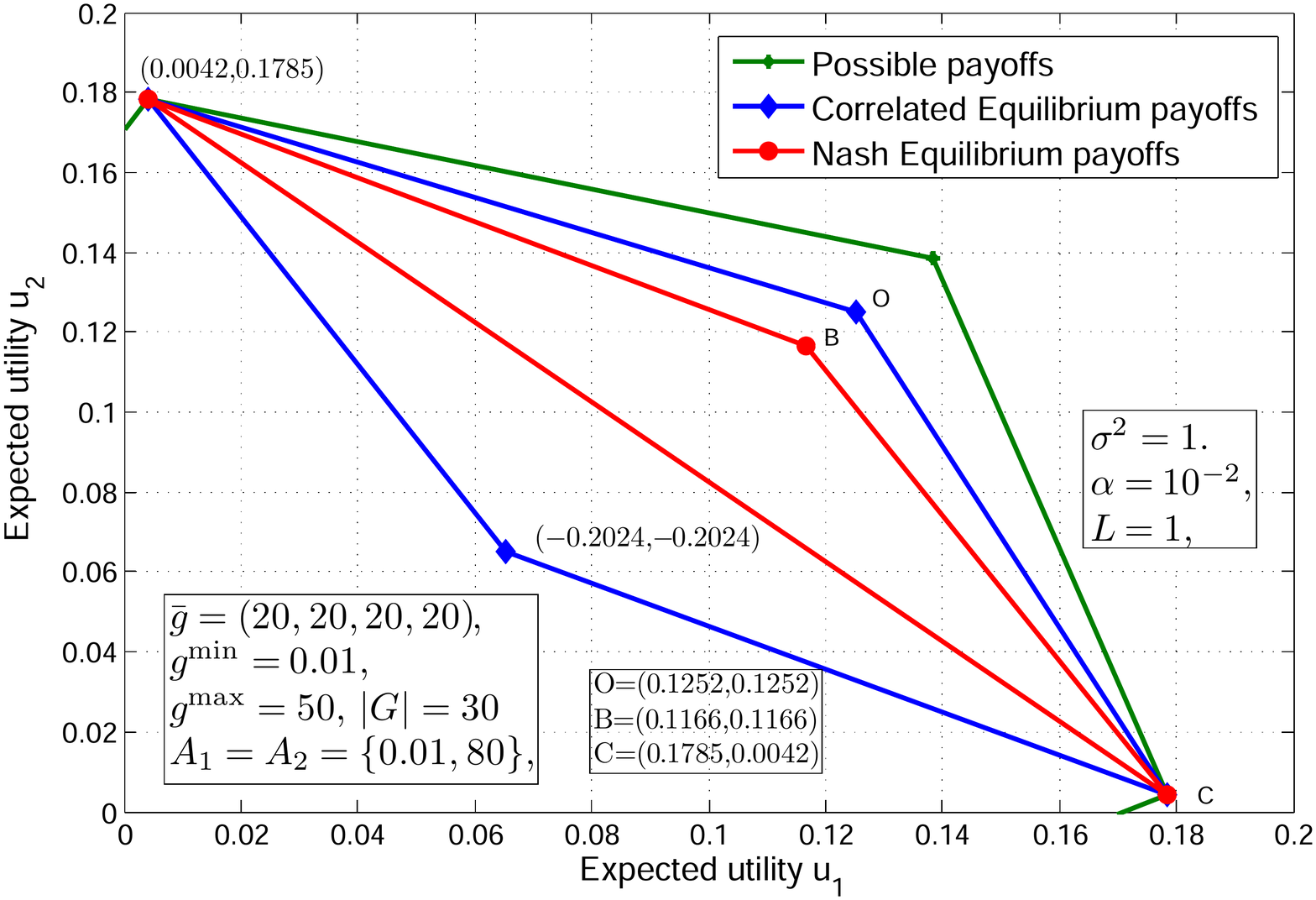}
		\caption{Payoffs of a two player power control game. Setting 1.} \label{fig:phase-II-ESNR-nb-bits-MEQ}
	\end{subfigure}
\hspace{0.01\linewidth}
	\begin{subfigure}[t]{.48 \linewidth}
		\centering
		 \includegraphics[width=7cm,height=5cm]{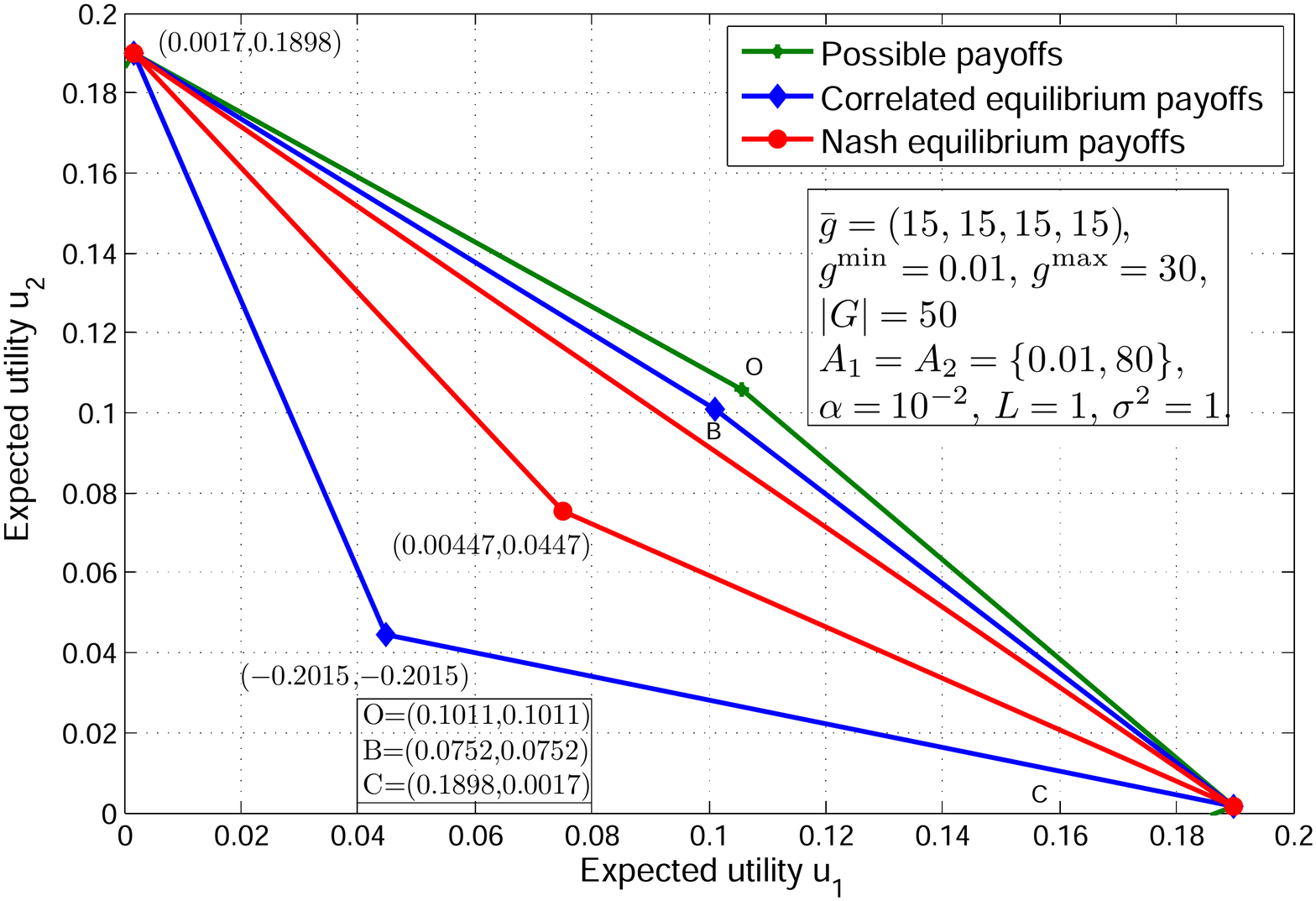}
		\caption{Payoffs of a two player power control game. Setting 2.}  \label{fig:phase-II-ESNR-nb-time-slots}
	\end{subfigure}
	\caption{Sets of: possible payoffs, correlated equilibrium payoffs and Nash equilibrium payoffs for two settings of parameters.}
\end{figure}

\begin{figure}[h]
	\centering
	\begin{subfigure}[t]{.48 \linewidth}
		\centering
		 \includegraphics[width=7cm,height=5cm]{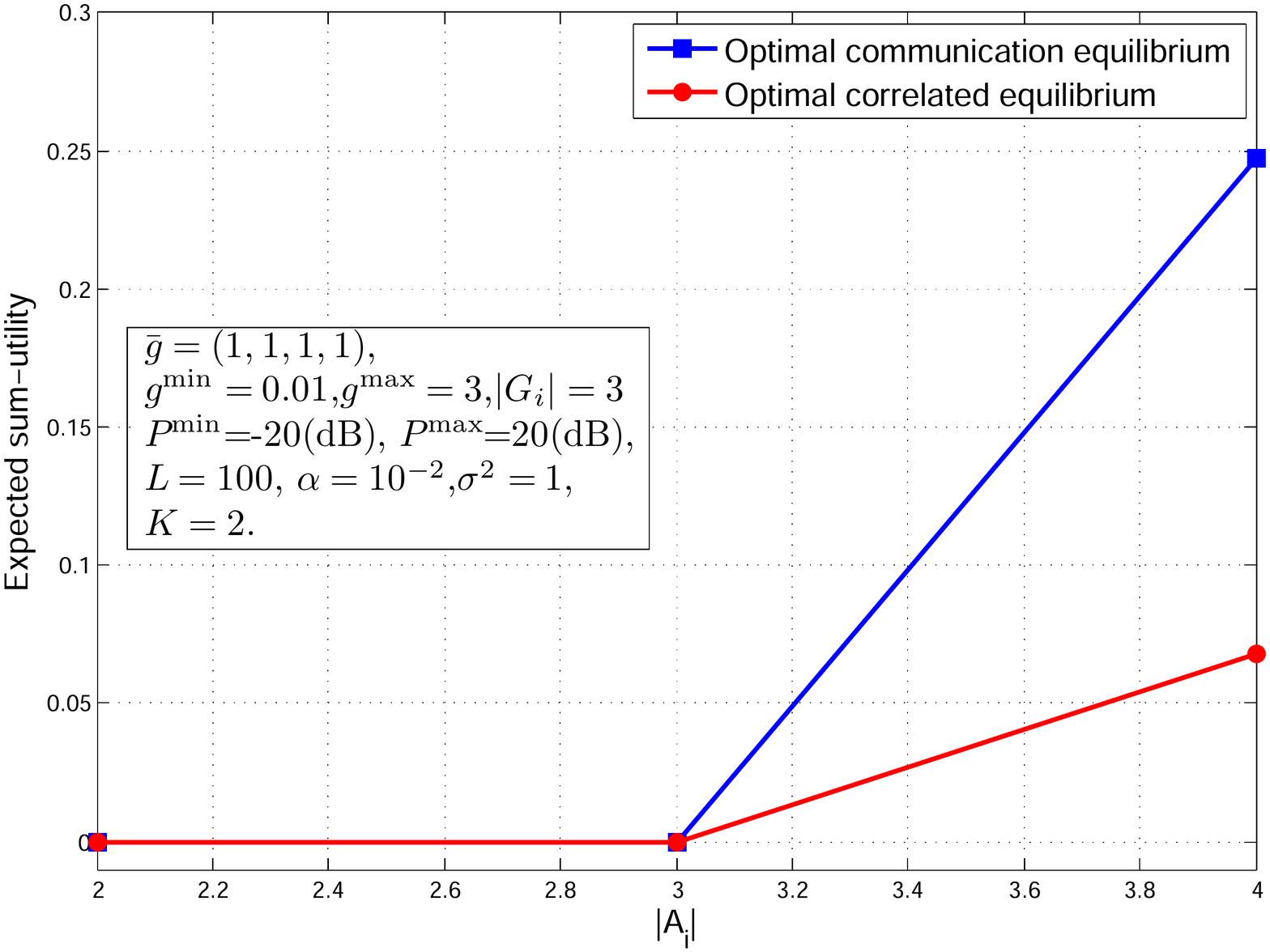}
		\caption{Expected sum-utility of correlated and communication equilibria against different number of actions $|A_i|$. Larger action sets are not plotted due to complexity issues.} \label{fig:phase-II-ESNR-nb-bits-MEQ}
	\end{subfigure}
\hspace{0.01\linewidth}
	\begin{subfigure}[t]{.48 \linewidth}
		\centering
		 \includegraphics[width=7cm,height=5cm]{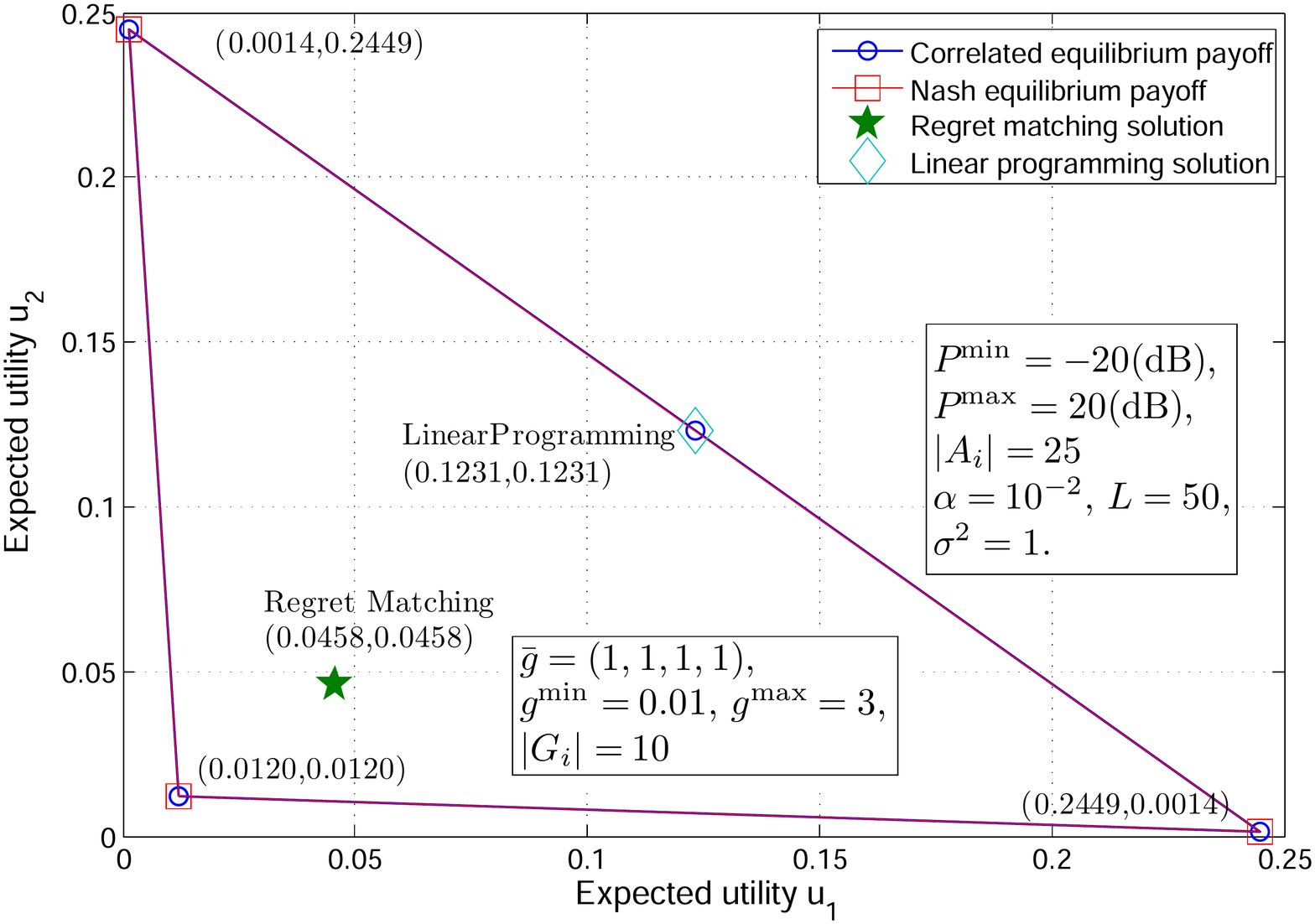}
		\caption{Sets of correlated equilibrium payoffs and Nash equilibrium payoffs. The correlated equilibrium provided by regret matching is also indicated.}  \label{fig:phase-II-ESNR-nb-time-slots}
	\end{subfigure}
\caption{Equilibrium payoffs for larger action sets.}
\end{figure}

Fig 1 compares the payoffs attained by the regular Nash equilibrium and the correlated equilibrium with the region of all possible payoffs (in a centralized setting). We observe that the correlated equilibrium can improve the utility of both players when compared to the Nash. We are limited to the two action case, but this is a very relevant case as seen from literature \cite{isit2013}. Fig 2a plots the utility of the correlated equilibrium and the communication equilibrium for higher action sets, but we plot to a maximum of four actions due to complexity issues in evaluating the communication equilibrium. Fig 2b compares the solution of regret matching with that of correlated equilibrium when the action sets are of size $25$. Fig 2b demonstrates that although the regret matching algorithm is computationally fast, the solutions are not Pareto-optimal.

% Fig. \ref{fig:phase-II-ESNR-nb-bits-MEQ} represents the different payoff sets of the power game presented in this paper. Three sets are represented, the largest set corresponds to the possible payoffs of the game, the middle set is the correlated equilibrium payoffs, finally, the shortest set corresponds to the Nash equilibrium payoffs. %The correlated equilibrium payoffs and Nash equilibrium payoffs set have been obtained by applying the procedure presented in Section \ref{sec:61}.
% The Fig. \ref{fig:phase-II-ESNR-nb-bits-MEQ} shows the contribution of the correlation among players' actions in payoffs. For the assumed setup, introducing correlation provides more important payoff set, compared to the set provided by the initial power control game without any correlation mechanism. As we can see, the set of Nash equilibrium payoffs is inside of the correlated equilibrium payoffs.

\section{Conclusion}

The goal of this paper is to make some progress in terms of knowing how communication and correlated equilibria can be used to achieve good tradeoffs between distributedness (in terms of observation and decision-wise) and global efficiency in power control problems, and more specifically when the utility is taken to be the goodput minus the transmit cost. Interestingly, our simulation results show encouraging results. However, an important challenge is left open, which is to know how to reach an efficient correlated equilibrium with a regret-matching-type learning algorithm.

%\newpage
%%%%%%%%%%%%%%%%%%%%%%%% referenc.tex %%%%%%%%%%%%%%%%%%%%%%%%%%%%%%
% sample references
%
%
% Use this file as a template for your own input.
%
%%%%%%%%%%%%%%%%%%%%%%%% Springer-Verlag %%%%%%%%%%%%%%%%%%%%%%%%%%

%
% BibTeX users please use
% \bibliographystyle{}
% \bibliography{}

%%%%%%%%%%%%%%%%%%%%%%%%%%%%%%%%%%%%%%%%%%%%%%%%%%%%%%%%%%%%%%%%%%%%%%

%%%%%%%%%%%%%%%%%%%%%%%%%%%%%%%%%%%%%%%%%%%%%%%%%%%%%%%%%%%%%%%%%%%%%%

\printindex
\end{document}